\documentclass[12pt,preprint]{aastex}

\slugcomment{The Astrophysical Journal, Submitted 2003 February 13}

\shortauthors{Wanajo et al.}
\shorttitle{Nucleosynthesis in a Collapsing O-Ne-Mg Core}

\begin{document}

\title{THE $r$-PROCESS IN SUPERNOVA EXPLOSIONS FROM THE COLLAPSE OF
O-Ne-Mg CORES}

\author{\sc Shinya Wanajo, Masaya Tamamura, and Naoki Itoh}
\affil{Department of Physics, Sophia University,
       7-1 Kioi-cho, Chiyoda-ku, Tokyo, 102-8554, Japan;\\
       wanajo@sophia.ac.jp, m-tamamu@sophia.ac.jp, n\_itoh@sophia.ac.jp}
\author{\sc Ken'ichi Nomoto}
\affil{Department of Astronomy, School of Science, University of Tokyo,
       Bunkyo-ku, Tokyo, 113-0033, Japan; nomoto@astron.s.u-tokyo.ac.jp}
\author{\sc Yuhri Ishimaru}
\affil{Department of Physics and Graduate School of Humanities and Sciences,
       Ochanomizu University, 2-1-1 Otsuka, Bunkyo-ku, Tokyo 112-8610, Japan;\\
       ishimaru@phys.ocha.ac.jp}
\author{\sc Timothy C. Beers}
\affil{Department of Physics/Astronomy, Michigan State University,
       E. Lansing, MI 48824, USA; beers@pa.msu.edu}

\and

\author{\sc Satoshi Nozawa}
\affil{Josai Junior College for Women, 1-1 Keyakidai, Sakado-shi,
       Saitama, 350-0290, Japan; snozawa@venus.josai.ac.jp}

\bigskip
\affil{The Astrophysical Journal, Submitted 2003 February 13}

\begin{abstract}

While the origin of $r$-process nuclei remains a long-standing mystery,
recent spectroscopic studies of extremely metal-poor stars in the
Galactic halo strongly suggest that it is associated with core-collapse
supernovae. In this study we examine $r$-process nucleosynthesis in a
``prompt supernova explosion'' from an $8-10 M_\odot$ progenitor star,
as an alternative scenario to the ``neutrino wind'' mechanism, which has
also been considered to be a promising site of the $r$-process. In the
present model, the progenitor star has formed an oxygen-neon-magnesium
core (of mass $1.38 M_\odot$) at its center. Its smaller gravitational
potential, as well as the smaller core that is in nuclear statistical
equilibrium at the time of core bounce, as compared to the iron cores in
more massive stars, may allow the star to explode hydrodynamically,
rather than by delayed neutrino heating.  The core-collapse simulations
are performed with a one-dimension, Newtonian hydrodynamic code.  We
obtain a very weak prompt explosion, in which no $r$-processing
occurs. We further simulate energetic prompt explosions by enhancement
of the shock-heating energy, in order to investigate conditions necessary
for the production of $r$-process nuclei in such events. The $r$-process
nucleosynthesis is calculated using a nuclear reaction network code
including relevant neutron-rich isotopes with reactions among them. The
highly neutronized ejecta ($Y_e \approx 0.14-0.20$) leads to robust
production of $r$-process nuclei; their relative abundances are in
excellent agreement with the solar $r$-process pattern. Our results
suggest that prompt explosions of $8-10 M_\odot$ stars with
oxygen-neon-magnesium cores can be a promising site of $r$-process
nuclei. The mass of the $r$-process material per event is about two
orders of magnitude larger than that expected from Galactic chemical
evolution studies. We propose, therefore, that only a small fraction of
$r$-process material is ejected, owing to the ``mixing-fallback''
mechanism of the core matter, wherein most of the $r$-process material
falls back onto the proto-neutron star.

A lower limit on the age of the universe is derived by application of the U-Th
chronometer pair by comparison with the observed ratio of these species in the
highly $r$-process enhanced, extremely metal-poor star CS~31082-001. The
inferred age is $14.1 \pm 2.4$~Gyr -- the same as that obtained previously
based on the neutrino wind scenario with the same nuclear mass formula. This
suggests that chronometric estimates obtained using the U-Th pair are
independent of the astrophysical conditions considered.

\end{abstract}


\section{INTRODUCTION}

The astrophysical origin of the rapid neutron-capture ($r$-process)
species has been a long-standing mystery. Recently, however, a number of
important new clues have been provided by spectroscopic studies of
extremely-metal-poor stars in the Galaxy. The appearance of
neutron-capture elements in these oldest stars in the Galaxy, including
the pure-$r$-process origin of elements such as thorium and uranium,
strongly suggests that the $r$-process nuclei have come from
core-collapse supernovae \citep{Sned96, Sned00, Sned03b, Cayr01, Cowa02,
Hill02}. \citet{Ishi99} have shown that the large star-to-star
dispersion of the observed abundances of neutron-capture elements
relative to iron in very metal-poor stars is also naturally explained if
the $r$-process elements originate from a limited mass range of
core-collapse supernovae with little iron production ($8-10 M_\odot$ or
$\ge 30 M_\odot$). Qian \& Wasserburg (2003; see also Wasserburg \& Qian
2000; Qian \& Wasserburg 2001, 2002) have proposed that the production
of the heavy $r$-process nuclei ($A > 130$) is decoupled from the
production of iron-peak and $\alpha$ nuclei by comparison of observed
abundances among the extremely metal-poor stars. In this view, the
production site of the heavy $r$-process nuclei is associated with the
accretion-induced collapse (AIC) of a (carbon-oxygen or
oxygen-neon-magnesium) white dwarf in a binary system \citep{Nomo91}, or
Type~II supernovae from $8-10 M_\odot$ stars \citep{Nomo84}.

So far, the ``neutrino wind'' scenario, in which the free nucleons
accelerated by the intense neutrino flux near the neutrino sphere of a
core-collapse supernova assemble to heavier nuclei, has been believed to
be the most promising astrophysical site of the $r$-process
\citep{Woos94}. Even this scenario, however, encounters some
difficulties \citep{Qian96, Hoff97, Card97, Otsu00, Thom01, Wana01,
Wana02}. For example, \citet{Wana01} have shown that an extremely
compact proto-neutron star, e.g., $2.0 M_\odot$ and 10~km, must be
formed in order to account for the solar $r$-process pattern, at least
within the framework of a spherically symmetric explosion.  Although its
possibility cannot be excluded, such a compact remnant is allowed by
only a few of the many existing equations of state of nuclear matter.

In addition, recent spectroscopic studies of extremely metal-poor stars
in the Galactic halo indicate that the observed abundance patterns of
the lighter ($Z < 56$) and heavier ($Z > 56$) neutron-capture elements
cannot be explained by a single astrophysical site (e.g., neutrino
winds); there must exist at least two different $r$-process sites
\citep{Ishi00, Qian01, John02, Sned03a}. Hence, it is of special
importance to consider alternative possibilities for the occurrence of
the $r$-process in core-collapse supernovae.
   
\citet{Nomo84, Nomo87} has shown that $8-10 M_\odot$ stars form an
electron-degenerate oxygen-neon-magnesium (O-Ne-Mg) core that does not undergo
further nuclear burning; rather, it directly collapses due to electron capture
on $^{24}$Mg and $^{20}$Ne (Miyaji et al. 1980). \citet{Hill84} have
demonstrated that the collapsing O-Ne-Mg core explodes in a prompt manner, and
\citet{Whee98} have suggested that the exploding O-Ne-Mg core could be a viable
site for the $r$-process. It has been pointed out that, if the core exploded
hydrodynamically prior to the onset of delayed neutrino heating (i.e., it
underwent a prompt explosion), the electron fraction (electron number per
baryon), $Y_e$, in the innermost layer of the ejecta would approach $\sim 0.2$
\citep{Hill84}. Earlier works have in fact shown that a robust $r$-process
proceeds in such conditions \citep{Schr73, Sato74, Hill76}.

Recently, \citet{Sumi01} have demonstrated that the prompt explosion of an
$11 M_\odot$ star with an iron core might also be a promising site of
the $r$-process. They obtained a prompt explosion by an adiabatic
core-collapse calculation without inclusion of electron capture and
neutrino transport. \citet{Thom03} have shown, however, that no
explosion is obtained with such a progenitor \citep{Woos95}, when
including electron capture along with a detailed treatment of neutrino
transport.  Many previous works have suggested that even the lowest mass
of core-collapse supernovae ($\sim 10 M_\odot$), in which a relatively
smaller iron core is formed, may have difficulties in achieving a
hydrodynamic explosion \citep{Bowe82, Burr83, Burr85, Brue89a, Brue89b,
Baro90}.

On the other hand, the question of whether $8-10 M_\odot$ stars that
form O-Ne-Mg cores can explode hydrodynamically is still open. The
possibility that these stars explode promptly remains because of the
smaller iron core present at the onset of the core bounce, as well as
the smaller gravitational potential of their collapsing
cores. \citet{Hill84} have obtained a prompt explosion of a $9 M_\odot$
star with a $1.38 M_\odot$ O-Ne-Mg core, while others, using the same
progenitor, have not \citep{Burr85, Baro87a}.  \citet{Mayl88} obtained
an explosion, not by a prompt shock, but by late-time neutrino
heating. Similar results can be seen in the studies of AICs. Note that
an AIC is an analogous phenomenon to a collapsing O-Ne-Mg core resulted
from a single $8-10 M_\odot$ star, since both consist of electron
degenerate cores \citep{Nomo91}. \citet{Frye99} have obtained an
explosion by neutrino heating, while others have not \citep{Baro87b,
Woos92a}. The reason for these different outcomes is due, perhaps, to
the application of different equations of state for dense matter,
although other physical inputs may also have some influence
\citep{Frye99}. Thus, even if a star of $8-10 M_\odot$ exploded, it
would be difficult to derive, with confidence, the physical properties
as well as the mass of the ejected matter. Given this highly uncertain
situation it is necessary to examine the resulting $r$-process
nucleosynthesis in explosions obtained with different sets of input
physics.

The purpose of this study is to investigate conditions necessary for the
production of $r$-process nuclei obtained in purely hydrodynamical models of
prompt explosions of collapsing O-Ne-Mg cores, and to explore some of the
consequences if those conditions are met. The core collapse and the
subsequent core bounce are simulated by a one-dimensional hydrodynamic
code with Newtonian gravity (\S~2). For simplicity, neutrino transport
is not taken into account. Hence, we focus on only hydrodynamical
explosions just after core bounce, not on the delayed explosions
obtained via late-time neutrino heating. As seen in \S~2, the explosion
is marginal, and no $r$-processing is expected.  In order to obtain
$r$-processed material, we find it necessary to force the occurrence of
more energetic explosions. The energetic explosions are simulated by
artificial enhancements of the shock-heating energy, rather than by
application of different sets of input physics, for simplicity. The
$r$-process nucleosynthesis in these explosions is then calculated with
the use of a nuclear reaction network code (\S~3). The resulting
contribution of the $r$-process material created in these simulations to
the early chemical evolution of the Galaxy is discussed in \S~4. The
results of chronometric age dating, using the U-Th chronometer pair
based on our nucleosynthesis results, is discussed in \S~5. A summary
and conclusions follow in \S~6.

\section{PROMPT EXPLOSION}

A pre-supernova model of a $9M_\odot$ star is taken from \citet{Nomo84},
which forms a 1.38 $M_\odot$ O-Ne-Mg core near the end of its evolution
\citep[see also][]{Miya80, Miya87, Nomo87}. We link this core to a
one-dimensional implicit Lagrangian hydrodynamic code with Newtonian
gravity \citep{Bowe91}. This core is modeled with a finely zoned mesh of
200 mass shells ($2\times 10^{-2} M_\odot$ to $0.8 M_\odot$, $5\times
10^{-3} M_\odot$ to $1.3 M_\odot$, and $5\times 10^{-3}-1\times 10^{-7}
M_\odot$ to the edge of the core).

The equation of state of nuclear matter (EOS) is taken from
\citet{Shen98}, which is based on relativistic mean field theory. The
equation of state for the electron and positron gas includes arbitrary
relativistic pairs as well as arbitrary degeneracy. Electron and positron
capture on nuclei, as well as on free nucleons, are included, along with the use
of the up-to-date rates from \citet{Lang00}. The capture is suppressed above the
neutrino trapping density, taken to be $3\times 10^{11}$~g~cm$^{-3}$, since the
neutrino transport process is not taken into account in this study. This
simplification may be appropriate, since neutrinos at the early epoch of the
core bounce do not appear to make a significant correction to the entropy
compared to that obtained from shock heating as shown by \citet{Hill84}. The
delayed neutrino heating may not significantly modify the mass trajectories of
the outgoing matter either, since the bulk of the ejecta are lifted to $\sim
1000$~km during the first few 100~ms from the onset of the core bounce (as shown
below), which is far from the location of the neutrino sphere (a few tens of
kilometers). It is evident, however, that an accurate treatment of neutrino
transport will be needed to obtain accurate mass trajectories in future work.

Nuclear burning is implemented in a simplified manner. The composition of the
O-Ne-Mg core is held fixed until the temperature in each zone reaches the onset
of oxygen-burning, taken to be $2\times 10^9$~K \citep{Nomo87}, at which point
the matter is assumed to be instantaneously in nuclear statistical equilibrium
(NSE). The temperature is then calculated by including its nuclear energy
release. It should be noted that we find a weak $\alpha$-rich freezeout in the
subsequent post-processing nucleosynthesis calculations (\S~3), owing to the
entropy, $\sim 10 N_A k$, in the ejecta. This means that the outgoing ejecta are
not in perfect NSE, which is assumed in our hydrodynamic calculations. An
improvement that takes the non-NSE matter properly into account will be needed
to obtain more accurate trajectories.

We begin the hydrodynamical computations with this pre-supernova model,
which has a density of $4.4\times 10^{10}$~g~cm$^{-3}$ and temperature
of $1.3\times 10^{10}$~K at its center. The inner $0.1 M_\odot$ has
already burned to NSE. As a result, the central $Y_e$ is rather low,
0.37, owing to electron capture. The core bounce is initiated when $\sim
90$~ms has passed from the start of the calculation. At this time the
NSE core contains only $1.0 M_\odot$, which is significantly smaller
than the cases of collapsing iron cores ($\gtrsim 1.3 M_\odot$). The
central density is $2.2\times 10^{14}$~g~cm$^{-3}$, significantly lower
than that of \citet[slightly above $3\times
10^{14}$~g~cm$^{-3}$]{Hill84}, although the temperature ($= 2.1\times
10^{10}$~K) and $Y_e$ (= 0.34), are similar. This difference is perhaps
due to the use of a relatively stiff EOS in this study.

We find that a very weak explosion results, with an ejected mass of $0.008
M_\odot$ and an explosion energy of $2\times 10^{49}$~ergs (model Q0 in
Table~1). The time variations of the radius, temperature, and density of each
zone are displayed in Figure~1. The lowest $Y_e$ in the outgoing ejecta is
0.45, where no $r$-processing is expected given the entropy of $\sim 10 N_A k$.
This is in contrast to the very energetic explosions, with ejected masses of
$0.2 M_\odot$, explosion energies of $2\times 10^{51}$~ergs, and low $Y_e$ of
$\sim 0.2$ obtained by \citet{Hill84}. This might be a consequence of the lower
gravitational energy release owing to the EOS applied in this study.

In order to examine the possible operation of the $r$-process in the
explosion of this model, we artificially obtain explosions with typical
energies of $\sim 10^{51}$~ergs by application of a multiplicative
factor ($f_{\rm shock}$) to the shock-heating term in the energy
equation (models~Q3, Q5, and Q6 in Table~1). We take this simplified
approach in this study, since the main difference between our result and
that by \citet{Hill84} appear to be the lower central density in
ours. If the inner core reached a higher density at the time of core
bounce by applying, for example, a softer EOS, the matter would obtain
higher shock-heating energy. This is clearly not a self-consistent
approach, and a further study is needed to conclude whether such a
progenitor star explodes or not, taking into account a more accurate
treatment of neutrino transport, as well as with various sets of input
physics (like EOSs). It should be emphasized, however, that our purpose
in this paper is not to justify the prompt explosions of collapsing
O-Ne-Mg cores, but to investigate the conditions necessary for the
production of $r$-process nuclei from such an event {\em if it occurs}.

Table~1 lists the multiplicative factor applied to the shock-heating
term ($f_{\rm shock}$), explosion energy ($E_{\rm exp}$), ejected mass
($M_{\rm ej}$), and minimum $Y_e$ in the ejecta obtained for each model. 
Energetic explosions with $E_{\rm exp} > 10^{51}$~ergs are obtained for
$f_{\rm shock} \ge 1.5$ (models~Q5 and Q6), in which deeper neutronized
zones are ejected by the prompt shock, as can be seen in Figure~2
(model~Q6). This is in contrast to the weak explosions with $E_{\rm exp}
\le 10^{50}$~ergs (models~Q0 and Q3), in which only the surface of the
core blows off (Figure~1). Note that the remnant masses for models~Q5
and Q6 are $1.19 M_\odot$ and $0.94 M_\odot$, respectively, which are
significantly smaller than the typical neutron star mass of $\sim 1.4
M_\odot$. We consider it likely that a mass of $\sim 1.4 M_\odot$ is recovered
by fallback of the once-ejected matter, as discussed in \S~4.

In Figure~3 the electron fraction in the ejecta of each model is shown
as a function of the ejected mass point, $M_{\rm ej}$. For models~Q0 and
Q3, $Y_e$ decreases steeply with $M_{\rm ej}$, since the duration of
electron capturing is long, owing to the slowly expanding ejecta
(Figure~1). For models~Q5 and Q6, on the other hand, $Y_e$ decreases
gradually with $M_{\rm ej}$, owing to the fast expansion of the outgoing
ejecta. Nevertheless, the inner regions approach very low $Y_e$, 0.30
and 0.14 for models~Q5 and Q6, respectively, owing to their rather high
density ($\sim 10^{11}$~g~cm$^{-3}$) at the time of core bounce
(Figure~2).  Note that, for model~Q6, $Y_e$ increases again for $M_{\rm
ej} > 0.3 M_\odot$.  This is due to the fact that the positron capture
on free neutrons overcomes the electron capture on free protons when the
electron degeneracy becomes less effective in the high temperature
matter. The innermost, slowly outgoing region suffers from this
effect. This can be also seen in the results of \citet{Mayl88}, who
obtained a neutrino-powered (not {\em prompt}) explosion with the same
O-Ne-Mg core.

The minimum $Y_e$ ($\approx 0.40$) in \citet{Mayl88} is, however,
significantly higher than ours. This is mainly due to neutrino capture
on free nucleons, since the matter is driven by intense neutrino flux in
their simulation. $Y_e$ might increase further once nucleons begin to
assemble into $\alpha$ particles and heavy nuclei \citep[the
``$\alpha$-effect'';][]{Meye98}. In the case of the prompt explosions
considered here, however, these neutrino effects may not alter $Y_e$
significantly. The reason is that the bulk of the ejecta are lifted to
$\sim 1000$~km at the arrival of the delayed neutrinos (a few 100~ms
from the onset of the core bounce), at which the capture timescale of
neutrinos on free nucleons is no less than a few seconds \citep[see,
e.g., eq.~(1) in][]{Qian97}. The dynamical timescales of the outgoing
mass shells in model~Q6 at $T_9 = 1, 3, 5$, and 7 (after the core
bounce), defined by $\tau_{\rm dyn} = |\rho/(d\rho/dt)|$, are shown in
Figure~4. As can be seen, the dynamical timescale prior to the
$r$-process phase ($T_9 \gtrsim 3$) is significantly smaller than that
of neutrino interaction.

The trend of the $Y_e - M_{\rm ej}$ relation up to $M_{\rm ej} \sim 0.2
M_\odot$ is similar in models~Q5 and Q6, although it is inverted at
$M_{\rm ej} \sim 0.14 M_\odot$, owing to the slightly different
contribution of the positron and electron capture on free nucleons
(Figure~3). Hence, the $Y_e - M_{\rm ej}$ relation between the surface and the
innermost layer of the ejecta is expected to be similar to that of model~Q6, as
long as the explosion is sufficiently energetic ($\gtrsim 10^{51}$~ergs). In the
subsequent sections, therefore, we focus only on model~Q6, which is taken to be
representative of cases where $r$-process nucleosynthesis occurs. The ejected
mass, $M_{\rm ej}$, is thus taken to be a free parameter, instead of simulating
many other models by changing $f_{\rm shock}$. Note that the results by
\citet{Hill84} and \citet{Sumi01} are very similar to the cases with $M_{\rm
ej} \sim 0.2 - 0.3 M_\odot$ in model~Q6.

\section{THE $r$-PROCESS}

The yields of $r$-process nucleosynthesis species, adopting the model described
in \S~2 for the physical conditions, are obtained by application of an
extensive nuclear reaction network code.  The network consists of $\sim
3600$ species, all the way from single neutrons and protons up to the
fermium isotopes ($Z = 100$). We include all relevant reactions, i.e.,
$(n, \gamma)$, $(p,\gamma)$, $(\alpha, \gamma)$, $(p, n)$, $(\alpha,
p)$, $(\alpha, n)$, and their inverses.  Reaction rates are taken from
Thielemann (1995, private communication) for nuclei with $Z \le 46$ and
from \citet{Cowa91} for those with $Z \ge 47$. The latter used the mass
formula by \citet{Hilf76}. The three-body reaction $\alpha (\alpha n,
\gamma)^9$Be, which is of special importance as the bottleneck reaction
to heavier nuclei, is taken from the recent experimental data of
\citet{Utsu01}. The weak interactions, such as $\beta$-decay,
$\beta$-delayed neutron emission (up to three neutrons), and electron
capture are also included, although the latter is found to be
unimportant.

The $\alpha$-decay chains and spontaneous fission processes are taken
into account only after the freezeout of all other reactions, as in
\citet{Wana02}. For the latter, all nuclei with $A \ge 256$ are assumed
to decay by spontaneous fission only. The few known nuclei undergoing
spontaneous fission for $A < 256$ are also included, along with their
branching ratios. Neutron-induced and $\beta$-delayed fissions, as well
as the contribution of fission fragments to the lighter nuclei, are
neglected. Obviously, these treatments of the fission reactions are
oversimplified. Nevertheless, this may be acceptable, at least to first
order. We leave more accurate treatment of these matters to future
work.

Each calculation is started at $T_9 = 9$ (where $T_9 \equiv
T/10^9$~K). The initial composition is taken to be that of NSE with the
density and electron fraction at $T_9 = 9$, and consists mostly of
free nucleons and alpha particles. 

For reasons outlined in \S~2, we examine the nucleosynthesis in model~Q6 only,
in which robust $r$-processing is expected. The resulted abundances in
several representative Lagrangian mass shells are depicted in
Figure~5. The initial electron fractions are 0.29, 0.20, 0.18, 0.16,
0.14, and 0.20 for the corresponding zone numbers 83, 92, 95, 98, 105,
and 132, respectively. As can be seen, a robust $r$-processing is
possible only for $Y_e \lesssim 0.20$. In particular, a substantial
amount of thorium and uranium are produced only when $Y_e$ is less than
0.18.

The mass-integrated abundances from the surface (zone 1) to the zones
83, 92, 95, 98, 105, and 132 are compared with the solar $r$-process
abundances \citep{Kapp89} in Figure~6 (models~Q6a-f in Table~2). The
latter is scaled to match the height of the first ($A = 80$) and third
($A = 195$) peaks of the abundances in models~Q6a-b and Q6c-f,
respectively. The ejecta masses of these models are listed in
Table~2. The nucleosynthesis result in model~Q5 (not presented
here) is expected to be similar to that of model~Q6a, because of the
resemblance of the $Y_e - M_{\rm ej}$ profiles between these models
(Figure~3). As can be seen in Figure~6, a solar $r$-process pattern for
$A \gtrsim 130$ is naturally reproduced in models~Q6c-f, while
models~Q6a-b fail to reproduce the third abundance peak. This implies
that the region with $Y_e < 0.20$ must be ejected to account for
production of the third $r$-process peak. Furthermore, to account for
the solar level of thorium ($A = 232$) and uranium ($A = 235, 238$)
production, the region with rather low $Y_e$ ($< 0.18$) must be ejected. 
Note that large deficiencies of nuclei at $A \approx 115$ may be
supplied if the slower rates of neutron capture in this region are
adopted, as demonstrated by \citet{Gori97}.

We find that, for models~Q6c-e, the lighter $r$-process nuclei with
$A < 130$ are somewhat deficient compared to the solar $r$-process
pattern (Figure~6c-e). This trend can be also seen in the observational
abundance patterns of the highly $r$-process-enhanced, extremely metal-
poor stars CS~22892-052 \citep{Sned03b} and CS~31082-001
\citep{Hill02}. In model~Q6f, the deficiency is outstanding because of
large ejection of the low $Y_e$ matter (Figure~3). This is in contrast
to the previous results obtained for the neutrino wind scenario, which
significantly {\em overproduce} the nuclei with $A \approx 90$
\citep{Woos94, Wana01}. The nuclei with $A < 130$ can be supplied by
slightly less energetic explosions, like models~Q6a-b (Figures~5a-b). It
is also possible to consider that these lighter $r$-process nuclei
originate from ``neutrino winds'' in more massive supernovae ($> 10
M_\odot$). The nuclei with $A < 130$ can be produced naturally in
neutrino winds with a reasonable compactness of the proto-neutron star,
e.g., $1.4 M_\odot$ and 10~km \citep{Wana01}.

Figure~6 implies that the production of thorium and uranium differs from
model to model, even though the abundance pattern seems to be {\em
universal} between the second and third $r$-process peaks, as seen in
models~Q6c-f. This is in agreement with recent observational results suggesting
that the ratio Th/Eu may exhibit a star-to-star scatter, while the abundance
pattern between the second and third peaks is in good agreement with the
solar $r$-process pattern (Honda et al., in preparation). Thus, the use of Th/Eu as
a cosmochronometer should be regarded with caution, at least until 
the possible variations can be better quantified; U/Th might
be a far more reliable chronometer, as discussed further in \S~5.

It is interesting to note that the nucleosynthesis results obtained for
models~Q6c-e are in good agreement with that of the prompt explosion of
a $11 M_\odot$ star with an iron core by \citet{Sumi01}. Our results
seem, however, in better agreement with the solar $r$-process pattern,
in particular near the rare-earth peak ($A \approx 160$) and the third
peak ($A \approx 195$). This is a consequence of the slower ejection of
the innermost region in our results owing to the reduction of pressure
by electron capture, which is not taken into account in
\citet{Sumi01}. In particular, the dynamical timescale of outgoing
matter becomes significantly long during the epoch of the $r$-process
($T_9 \sim 3$ to 1), as can be seen in Figure~4. The extremely low $Y_e$
($< 0.2$) drives matter near the neutron-drip line, where the abundance
pattern deviates from the solar one. In our models, however, the solar
$r$-process pattern is recovered by the ``freezeout effect'' after the
epoch of $r$-processing, as discussed in detail by \citet{Wana02}. This
arises because the matter stays at higher temperatures for a
longer time, hence the quasi-equilibrium between neutron emission by
photodisintegration and subsequent neutron-capture processes continues
to operate \citep{Surm01}.

It is also interesting to note that the mass of the $r$-processed ejecta
in our results ($\gtrsim 0.05 M_\odot$) is more than one order of
magnitude larger than that obtained by \citet[$\sim 0.003
M_\odot$]{Sumi01}. This is also a consequence of the longer dynamical
timescale of outgoing ejecta in our simulation. The $\alpha$ particles
are mostly consumed prior to the $r$-process phase in our calculations,
whereas they dominate the final nucleosynthesis yields in
\citet{Sumi01}. These results strongly suggest that an accurate
treatment of electron capture, as well as of neutrino transport, will be
crucial to derive an accurate prediction of the $r$-process abundances
from prompt explosions of O-Ne-Mg cores.

\section{CONTRIBUTION TO CHEMICAL EVOLUTION OF THE GALAXY}

We have demonstrated in the previous section that, if stars of $8-10 M_\odot$
do explode energetically, the solar $r$-process pattern can be reproduced quite
naturally. It is of importance, however, to see if the nucleosynthetic material
contributed to the Galaxy from such stars is consistent with the currently
observed elemental abundances in the solar neighborhood.

Figure~7 shows the ``production factors'' per supernova for model~Q6e,
defined for each nuclide as the final mass fraction, $X_{\rm ej}$,
diluted by the total ejected mass ($7.9 M_\odot$) from the $9 M_\odot$
star, divided by its solar abundance $X_\odot$ \citep{Ande89}. The solid
lines connect isotopes of a given element (after nuclear decay). The
dotted horizontal lines indicate a ``normalization band'' between the
largest production factor ($^{129}$Xe) and that by a factor of ten less
than that, along with a median value (dashed line). This band is taken
to be representative of the uncertainty in the nuclear data for very
neutron-rich nuclei \citep{Woos94}. As can be seen in Figure~7, a
majority of the nuclei with $A > 100$ fall within the normalization band
(except for the large deficiencies near $A = 120$), which is regarded to
be the dominant species produced. The deficiencies near $A = 88, 138$,
and 208 may be supplied by $s$-process contributions from other sources. 
Although such a supernova might be considered a likely production
site of $r$-process nuclei, its contribution to nuclei with $A < 80$,
including the $\alpha$ elements and iron-peak species, is negligible.

One of the essential questions raised by previous works is that, if
prompt supernova explosions are one of the major sites of $r$-process
nuclei, would in fact the $r$-process nuclei be significantly {\it
overproduced} \citep{Hill76}. As far as the explosion is purely
hydrodynamical, a highly neutronized deeper region must be ejected in
order for a successful $r$-process to result. It seems inevitable,
therefore, that one must avoid an ejection of large amounts of
$r$-process matter, at least when assuming spherical symmetry. Our
result shows that more than $0.05 M_\odot$ of the $r$-process matter ($A
\ge 120$) is ejected per event, which reproduces the solar $r$-process
pattern (models~Q6c-f in Table~2). This is about three orders of
magnitude larger than the $5.8 \times 10^{-5} M_\odot$ in the
neutrino-heated supernova ejecta from a $20 M_\odot$ star obtained by
\citet{Woos94}. This value might would be reduced to some extent by
$\alpha$-rich freezeout in faster outgoing mass shells than ours as
observed in \citet{Sumi01}. However, good agreement with the solar
$r$-process pattern would not be achieved in the ejecta expanding too
fast, as discussed in \S~3.

As discussed by \citet{Woos94}, production factors must be on the order of
$\sim 10$ for the case that all supernovae contribute equally to $r$-process
production in the Galaxy. Stars of $8-10 M_\odot$ would account for $\sim 30
\%$ of all supernovae (if they explode). The ejected masses ($\sim 7-9
M_\odot$) are smaller than a factor of two to three than those from more
massive supernovae. Thus, production factors of $\sim 100$ are
allowed in the case that all $8-10 M_\odot$ stars contribute equally to the
Galactic $r$-process material. The production factors in our results
are, however, about three orders of magnitude higher than this (Figure~7).

It might be argued that this type of event is extremely rare,
accounting for only $0.01-0.1 \%$ of all core-collapse
supernovae. However, observations of extremely metal-poor stars (${\rm
[Fe/H]} \sim -3$) in the Galactic halo show that at least two,
CS~22892-052 and CS~31082-001, out of about a hundred studied at high
resolution, imply contributions from highly $r$-process-enhanced
supernova ejecta \citep{Sned00, Hill02}. Moreover, such an extremely
rare event would result in a much larger dispersion of $r$-process
elements relative to iron than is currently observed amongst extremely
metal-poor stars. \citet{Ishi99} demonstrated that the observed
star-to-star dispersion of ${\rm [Eu/Fe]}$ over a range $\sim -1 $ to
2 dex, was reproduced by their chemical evolution model if Eu originated from
stars of $8-10 M_\odot$. Recent abundance measurements of Eu in a few
extremely metal-poor stars with ${\rm [Fe/H]} \lesssim -3$ by SUBARU/HDS
further supports their result (Ishimaru et al., in preparation). The
requisite mass of Eu in their model is $\sim 10^{-6} M_\odot$ per event. 
The ejected mass of Eu in our result is more than two orders of
magnitude larger (Table~2).

In order to resolve this conflict, we propose that the
``mixing-fallback'' mechanism operates in this kind of supernova. The
peculiar abundance patterns of some extremely iron-deficient stars,
including HE~0107-5240 with $[{\rm Fe/H}] = - 5.3$ \citep{Chri02}, is
explained successfully with this mechanism, as proposed by
\citet{Umed02, Umed03}. If a substantial amount of the hydrogen and
helium envelope remains at the onset of the explosion, the outgoing
ejecta may undergo large-scale mixing by Rayleigh-Taylor instabilities,
which is believed to have happened in SN1987A \citep{Hach90, Hera94,
Kifo03}. Thus a tiny amount, say, $\sim 1 \%$, of the $r$-process
material is mixed into the outer layers and then ejected, but most of
the core material may fall back onto the proto-neutron star via the
reverse shock arising from the hydrogen-helium layer interface. In this
case, the typical mass of the proto-neutron star ($\sim 1.4 M_\odot$) is
recovered. An asymmetric explosion mechanism, such as that which might
arise from rotating cores, may have a similar effect as the ejection of
deep-interior material in a small amount \citep{Yama94, Frye00}. This
``mixing-fallback'' scenario must be further tested by detailed
multidimensional-hydrodynamic studies.  However, it may provide us with
a new paradigm for the nature of supernova nucleosynthesis.

This type of event may be characterized as a ``faint'' supernova, owing
to the weakened explosion energy as well as the reduced amount of
$^{56}$Ni by ``mixing-fallback''. In addition, its ejecta consists
mostly of hydrogen and helium with a higher ratio of He/H than that in
the solar system. This event can be easily distinguished from the
core-collapse supernovae of iron cores resulted from stars of $> 10
M_\odot$, which are characterized by abundant $\alpha$-elements. It is
interesting to note that such an explosion bears a close resemblance to
the Crab supernova, SN1054 \citep{Nomo82}. There is the possibility,
therefore, that direct evidence will be obtained from the Crab nebula
(or other similar supernova remnants) through the detection of
$r$-process elements by future observations \citep{Wall95}. Detection of
the $\gamma$-ray lines from decays of radioactive isotopes produced by
the $r$-process could provide additional direct evidence for this scenario
\citep{Qian98}.

It is of special importance to confirm, from spectroscopic studies of
extremely metal-poor stars in the Galactic halo, that the $r$-process
elements are {\it not} associated with the production of $\alpha$- and
iron-peak elements \citep{Wass00, Qian01, Qian02}. \citet{Qian03} have
suggested that the $r$-process enrichment in extremely metal-poor stars
HD~115444, HD~122563, and CS~31082-001 is independent of the production
of the elements from Na to Zn (including $\alpha$- and iron-peak
elements). The abundances of Na-Zn among these stars are mostly the
same, while the level of $r$-process enhancement differs from star to
star. These authors take this as evidence that the heavy
$r$-process nuclei originated from AICs or Type~II supernovae from $8-10
M_\odot$ stars, although they prefer AICs for the $r$-process site. Note
that the nucleosynthetic outcome of an AIC event may be similar to the
collapsing O-Ne-Mg core resulting from a single $8-10 M_\odot$
star. Hence, it is possible that AICs also undergo prompt
explosions. The frequency of the AIC events in the Galaxy is expected to
be very small, $\sim 10^{-5}$~yr$^{-1}$ \citep{Bail90}. Assuming that
the frequency of the core-collapse supernovae in the Galaxy is $\sim
10^{-2}$~yr$^{-1}$ \citep{Capp97}, the production factor per AIC event
is $\sim 10^4$. This is in reasonable agreement with our result of
$\approx 10^5$, as can be seen in Figure~7. In the case of AICs,
however, there is no chance to undergo ``mixing-fallback'', owing to the
absence of the outer envelope. Thus, a much larger dispersion of [Eu/Fe]
than that observed in the extremely metal-poor stars seems difficult
to avoid.

We consider, therefore, that the heavy $r$-process nuclei originate from $8-10
M_\odot$ stars discussed in this study, being independent of whether they are
single stars or are in binary systems. The lighter $r$-process elements ($Z <
56$) observed in extremely metal-poor stars may represent simply the
interstellar medium from which these stars were formed, which originated from,
perhaps, ``neutrino winds'' in supernovae from stars of $> 10 M_\odot$. The
observed abundances of Eu relative to iron in extremely metal-poor stars are in
fact well reproduced with this assumption in the chemical evolution model by
\citet{Ishi99}. Future spectroscopic studies of extremely metal-poor stars may
be able to distinguish the scenario suggested in this study from AICs, for
example, through observations of $s$-process abundances originating from mass
transfer in binary systems.

\section{U-Th COSMOCHRONOLOGY}

The recently discovered $r$-process-enhanced, extremely metal-poor star,
CS~31082-001 (${\rm [Fe/H]} = -2.9$) has provided a new, potentially
quite powerful cosmochronometer, uranium
\citep{Cayr01,Hill02}. \citet{Wana02} have determined the age of this
star (more precisely, the time that has passed since the production
event of the $r$-process species incorporated into this star) using the
U-Th chronometer pair to be $14.1 \pm 2.5$~Gyr, based on the neutrino
wind scenario. This is regarded as a hard lower limit of the age of the
universe, and is in good agreement with that derived by an
site-independent approach \citep{Gori01,Scha02}. It is of interest to
examine if the same age of this star is also obtained, based on the
prompt explosion model of a $9 M_\odot$ star described in the previous
sections. Note that the same mass formula \citep{Hilf76} as in
\citet{Wana02} is used in this study for comparison.

In Figure~8 the available spectroscopic abundance data for CS~31082-001
(dots) are compared with the nucleosynthesis result of model~Q6e
discussed in \S~3 (thick line) and with the solar $r$-process pattern
(thin line), scaled at Eu ($Z = 63$). The data for the neutron-capture
elements in this star are taken from \citet{Hill02}. An overall
agreement of our result with the spectroscopic data up to lead ($Z =
82$) can be seen, although our result appears somewhat deficient for the
lighter elements.  This is in contrast to the results in neutrino winds
\citep{Wana02}, in which the lighter elements are significantly
overproduced. Thus, this model might be a reasonable one for the
$r$-process events that produce large amounts of thorium and uranium,
and whose products now appear in the atmosphere of CS~31082-001.

Figure~9 shows the mass-integrated abundance ratios of Th/Eu (open
squares) and U/Th (open circles) from the surface to the mass point
$M_{\rm ej}$, as well as the inferred ages of CS~31082-001, $t^* ({\rm Th/Eu})$
(filled squares) and $t^* ({\rm U/Th})$ (filled circles). As discussed
by \citet{Hill02}, the age of CS~31082-001 can be inferred by
application of the following relations:
\begin{eqnarray}
t^* ({\rm Th/Eu}) & = & 
46.67\, [\log ({\rm Th/Eu})_0 - \log ({\rm Th/Eu})_{\rm now}]\ {\rm Gyr}
\\
t^* ({{\rm U/Th}}) & = & 
21.76\, [\log ({\rm U/Th})_0 - \log ({\rm U/Th})_{\rm now}]\ {\rm Gyr}
\end{eqnarray}
where the half lives of $^{232}$Th (14.05 Gyr) and $^{238}$U (4.468 Gyr),
and the subscripts ``0'' and ``now'' denote the initial and current
values derived by theory and observation, respectively. In principle,
$t^* ({{\rm U/Th}})$ may serve as a more precise chronometer than $t^* ({{\rm
Th/Eu}})$, owing to the smaller coefficients in front of
equations~(2). Moreover, the ratio U/Th is less dependent on the model
parameter $M_{\rm ej}$, since these species are separated by only two
units in atomic number. Note that $^{235}$U is assumed to have
$\alpha$-decayed away because of its relatively short half life (0.704
Gyr).

As can be seen in Figure~9, U/Th approaches a constant value ($= 0.51$) for
$M_{\rm ej} \gtrsim 0.26 M_\odot$ ($Y_e \lesssim 0.17$), while Th/Eu varies
widely. As a result, the age of CS~31082-001 determined by U/Th results in a
constant value, $t^* ({{\rm U/Th}}) = 14.1$~Gyr, for $M_{\rm ej} \gtrsim 0.26
M_\odot$. The age $t^* ({{\rm Th/Eu}})$ is sensitive to the parameter $M_{\rm
ej}$, ranging from a negative age to 23.8~Gyr, which illustrates that caution
must be used in the application of this chronometer pair. It is useful, however,
to take $t^* ({{\rm Th/Eu}})$ as a constraint on the model parameter $M_{\rm
ej}$, although the result might be changed if one includes an accurate neutrino
transport and other input physics. It is found that the models with $M_{\rm ej}
= 0.30 M_\odot$ and $0.37 M_\odot$ (zone numbers 105 and 118, respectively) give
the same ages between $t^* ({{\rm Th/Eu}})$ and $t^* ({{\rm U/Th}})$. The former
corresponds to model~Q6e, which would provide a consistent scenario for the
origin of the $r$-process elements in CS~31082-001, as can be seen in Figure~8.
It should be noted that the fission fragments in model~Q6e account for 23\% of
the mass contained in $A \ge 120$ nuclei (the last column in Table~2), whose
contribution to the abundance pattern is neglected in this study. Obviously,
more accurate treatment of fission reactions is needed. Nevertheless, the age
$t^* ({{\rm U/Th}})$ would not be altered significantly by this improvement,
since the ratio U/Th is at the saturated value ($= 0.51$) at $M_{\rm ej} = 0.26
M_\odot$, where the mass fraction of fission fragments is only 2\%.

It is noteworthy that the inferred age by U/Th in this study is the same
as the result in \citet{Wana02}, which is based on a different
astrophysical scenario (neutrino winds), but uses the same nuclear mass
formula \citep{Hilf76}. Our result confirms the robustness of the age
determination using the U-Th chronometer pair, which is mostly
independent of the astrophysical conditions considered. Rather than the
$r$-process site, the nuclear mass formulae adopted, as well as the
treatment of fission are crucial as far as the U-Th pair is concerned
\citep{Seeg70, Gori01, Scha02}.

\section{SUMMARY AND CONCLUSIONS}

We have examined the $r$-process nucleosynthesis obtained in the prompt explosion
arising from the collapse of a $9 M_\odot$ star with an O-Ne-Mg
core. The core collapse and subsequent core bounce were simulated
with a one-dimensional, implicit, Lagrangian hydrodynamic code with
Newtonian gravity. Neutrino transport was neglected for simplicity. We
obtained a very weak explosion (model~Q0) with an explosion energy of
$\sim 2 \times 10^{49}$~ergs, and an ejected mass of $\sim 0.008
M_\odot$. No $r$-processing occurred in this model, because of the high
electron fraction ($\gtrsim 0.45$) with low entropy ($\sim 10 N_A k$).

We further simulated energetic explosions by an artificial enhancement
of the shock-heating energy, which might be expected from calculations
with other sets of input physics, as well as with other pre-supernova
models. This resulted in an explosion energy of $\gtrsim 10^{51}$~ergs
and an ejected mass of $\gtrsim 0.2 M_\odot$.  Highly neutronized matter
($Y_e \approx 0.14$) was ejected, which led to strong $r$-processing
(model~Q6).  Material arising from $r$-process nucleosynthesis was
calculated with a nuclear reaction network code containing $\sim 3600$
isotopes with all relevant reactions. The result was in good agreement
with the solar $r$-process pattern, in particular for nuclei with $A >
130$. Some of lighter $r$-process nuclei ($A < 130$) were deficient,
which is consistent with the abundance patterns of the highly $r$-process
enhanced, extremely metal-poor stars CS~22892-052 and
CS~31082-001. This implies that the lighter $r$-process nuclei may
originate from another site, which we suggest might be associated with
the ``neutrino wind'' in core-collapsing supernovae of iron cores
resulted from more massive stars ($> 10 M_\odot$).

The large ejection of $r$-process material ($\gtrsim 0.05 M_\odot$ per
event) conflicts with the level of dispersion of $r$-process elements
relative to iron observed in extremely metal-poor stars. We suggest,
therefore, that only a small fraction ($\sim 1\%$) of the $r$-processed
material is ejected, while the bulk of such material falls back onto the
proto-neutron star by the ``mixing-fallback'' mechanism. 

The age of the highly $r$-process enhanced, extremely metal-poor star,
CS~31082-001 was derived by application of the U-Th chronometer pair,
and can be regarded as a hard lower limit on the age of the
universe. The age obtained is $14.1 \pm 2.4$~Gyr (the quoted error only
includes that arising from the observations), the same as that based on
a different astrophysical site, the neutrino-wind scenario
\citep{Wana02}, using the same nuclear mass formula. This confirms that
the age determined by the U-Th pair is mostly independent of the
astrophysical environment considered. The dependence of age dating on
different nuclear mass formulae, based on the prompt explosion scenario
presented in this paper, will be reported in future work.

It is obvious that more studies, including an accurate treatment of neutrino
transport and other input physics, as well as multi-dimensional simulations of
prompt supernova explosions, are needed to derive the final conclusion on the
$r$-process scenario presented in this study. Nevertheless, this scenario is
attractive as a promising site of the $r$-process, since the solar $r$-process
pattern can be naturally reproduced without the problematic overproduction of $A
\approx 90$ that appeared in the neutrino wind scenario. This type of event is
characterized with the absence of $\alpha$- and iron-peak elements, which can be
easily distinguished from that of the core-collapsing iron core resulted from a
more massive star ($> 10 M_\odot$). Future spectroscopic studies of extremely
metal-poor stars in the Galactic halo will reveal if the collapsing O-Ne-Mg
cores of $8-10 M_\odot$ stars are a viable site for the production of
$r$-process nuclei. Improved observational determination of the U/Th ratio in
CS~31082-001, as is presently being pursued, as well as a measured abundance of
Pb in this star (as is being obtained with the Hubble Space Telescope), and the
identification of a greater number of highly r-process-enhanced, metal-poor
stars, also underway, will surely deepen our understanding of the relevant
processes involved.

\acknowledgments

We would like to acknowledge K. Sumiyoshi for providing an EOS table of nuclear
matter applied in this study and for helpful discussions. We also thank G.
Martinez-Pinedo for providing a data table of weak rates. We also acknowledge
the contributions of an anonymous referee, which led to clarification of a
number of points in our original presentation. This work was supported by a
Grant-in-Aid for Scientific Research (13640245, 13740129, 14047206, 14540223)
from the Ministry of Education, Culture, Sports, Science, and Technology of
Japan. T.C.B. acknowledges partial support from grants AST 00-98508 and AST
00-98549 awarded by the U.S. National Science Foundation.

\clearpage

\begin{deluxetable}{lllll}
\tablecaption{Results of Core-Collapse Simulations \label{tab1}}
\tablewidth{0pt}
\tablehead{
\colhead{model} &
\colhead{$f_{\rm shock}$} &
\colhead{$E_{\rm exp}$ ($10^{51}$~ergs)} &
\colhead{$M_{\rm ej}$ ($M_\odot$)} &
\colhead{$Y_{e, {\rm min}}$}}
\startdata
   Q0 &   1.0 &   0.018  &   0.0079 &   0.45 \\
   Q3 &   1.3 &   0.10   &   0.029  &   0.36 \\
   Q5 &   1.5 &   1.2    &   0.19   &   0.30 \\
   Q6 &   1.6 &   3.5    &   0.44   &   0.14 \\
\enddata
\end{deluxetable}

\clearpage

\begin{deluxetable}{lllllll}
\tablecaption{Ejected Mass ($M_\odot$) \label{tab2}}
\tablewidth{0pt}
\tablehead{
\colhead{model} &
\colhead{$M_{\rm ej}$} &
\colhead{$A\ge 120$} &
\colhead{$^{56}$Ni} &
\colhead{Fe} &
\colhead{Eu} &
\colhead{fission\tablenotemark{\dag}}
}
\startdata
Q0  & 0.0079 & 0.0               & 0.0018 & 0.0019 & 0.0 & 0.0 \\
Q6a & 0.19 & $2.6\times 10^{-4}$ & 0.018  & 0.020  & 0.0 & 0.0 \\
Q6b & 0.24 & 0.035 & 0.018 & 0.020 & $2.4\times 10^{-4}$ & $1.4\times 10^{-6}$ \\
Q6c & 0.25 & 0.051 & 0.018 & 0.020 & $4.1\times 10^{-4}$ & $4.3\times 10^{-4}$ \\
Q6d & 0.27 & 0.064 & 0.018 & 0.020 & $4.3\times 10^{-4}$ & 0.047 \\
Q6e & 0.30 & 0.080 & 0.018 & 0.020 & $4.6\times 10^{-4}$ & 0.23 \\
Q6f & 0.44 & 0.21  & 0.018 & 0.020 & 0.0020 & 0.15 \\
\enddata
\tablenotetext{\dag}{mass fraction of fission fragments ($A \ge 120$)}
\end{deluxetable}

\clearpage

\begin{figure}
\epsscale{0.6} 
\plotone{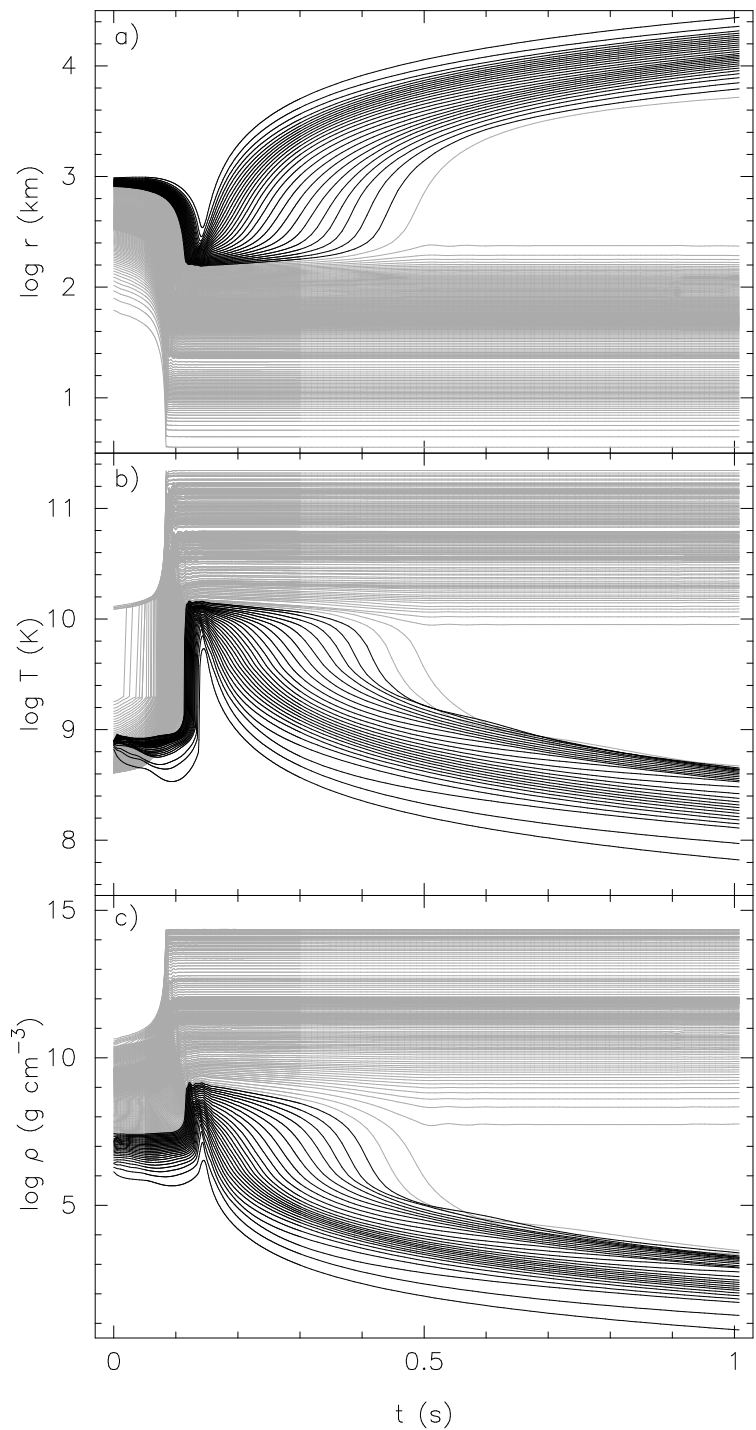} 
\caption{ Time variations of (a) radius,
(b) temperature, and (c) density for all mass points in the weak prompt
explosion of a $9 M_\odot$ star (model~Q0). The ejected mass points are
denoted in black, while those of the remnant are in grey.  }
\end{figure}

\clearpage

\begin{figure}
\epsscale{0.6} 
\plotone{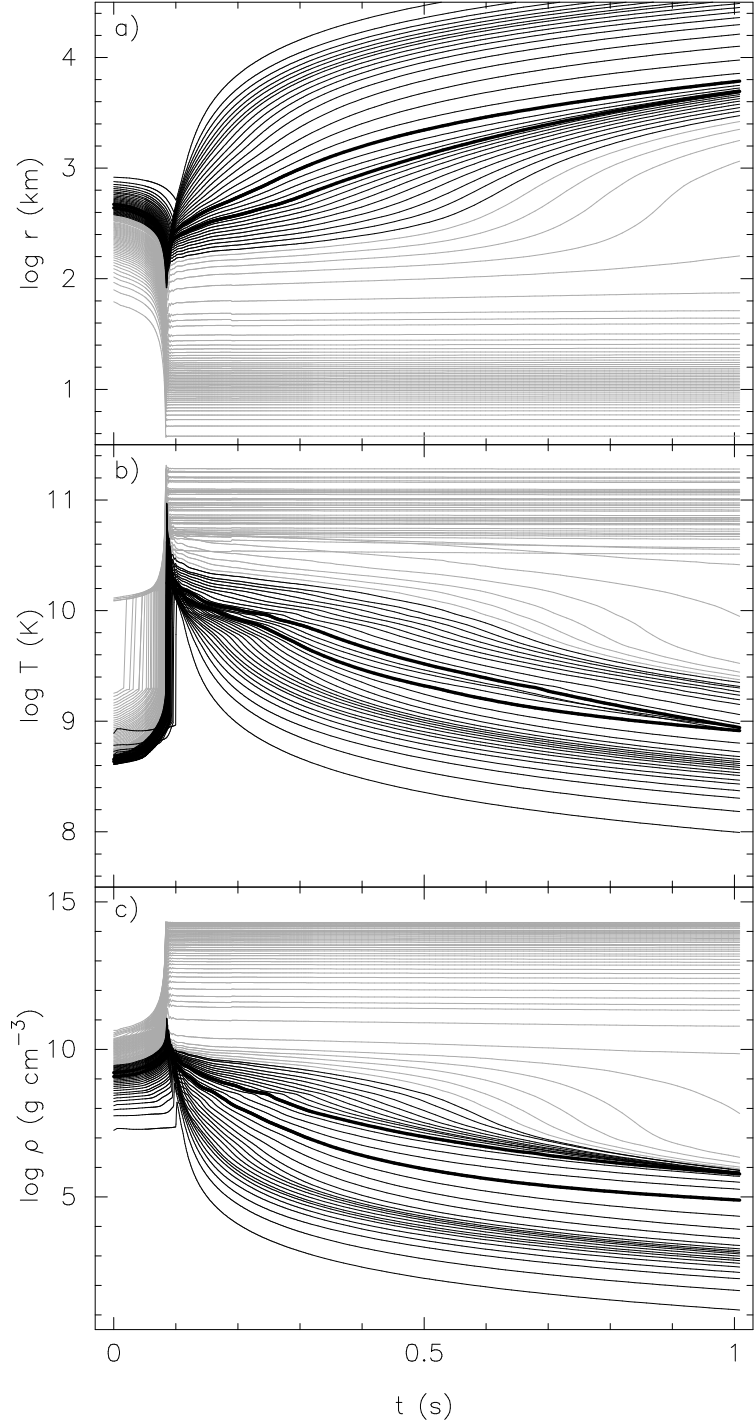} 
\caption{ Time variations of (a) radius,
(b) temperature, and (c) density for selected mass points (with roughly
an equal mass interval) in the energetic prompt explosion of a $9
M_\odot$ star, in which the shock-heating energy is enhanced
artificially by a factor of 1.6 (model~Q6). The ejected mass points are
denoted in black, while those of the remnant are in grey. Thick lines
denote the mass shells with $M_{\rm ej} = 0.2$ and $0.3 M_\odot$ ($Y_e =
0.20$ and 0.14, respectively; see Figure~3). The material between these
lines is particularly of importance to account for the solar $r$-process
abundances (see discussion in \S~3).}
\end{figure}

\clearpage

\begin{figure}
\epsscale{1.0} 
\plotone{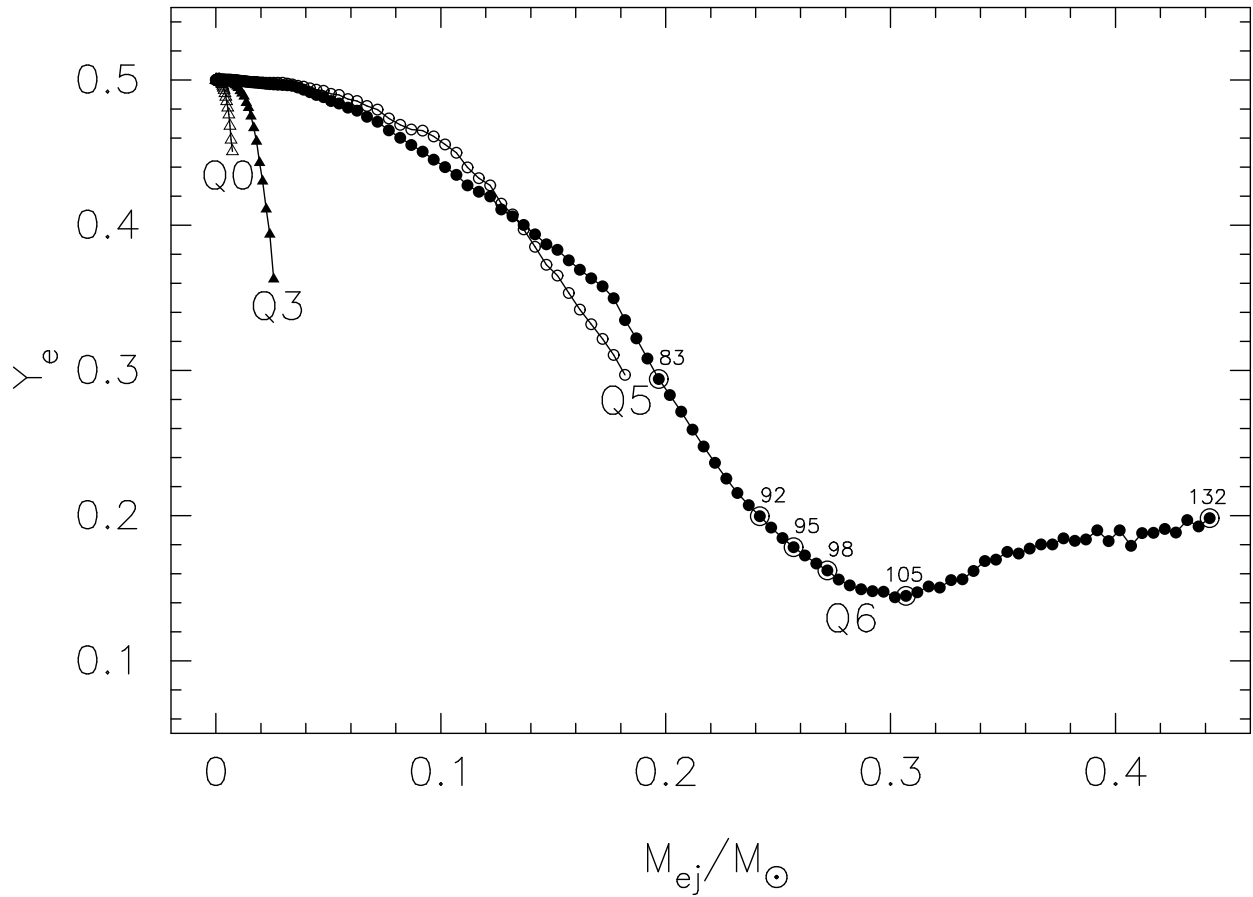} 
\caption{ $Y_e$ distribution in the
ejected material in models~Q0 (open triangles), Q3 (filled triangles),
Q5 (open circles), and Q6 (filled circles). The surface of the O-Ne-Mg
core is at mass coordinate zero. For model~Q6, selected mass points are
denoted by zone numbers (see Table~2).  }
\end{figure}

\clearpage

\begin{figure}
\epsscale{1.0} 
\plotone{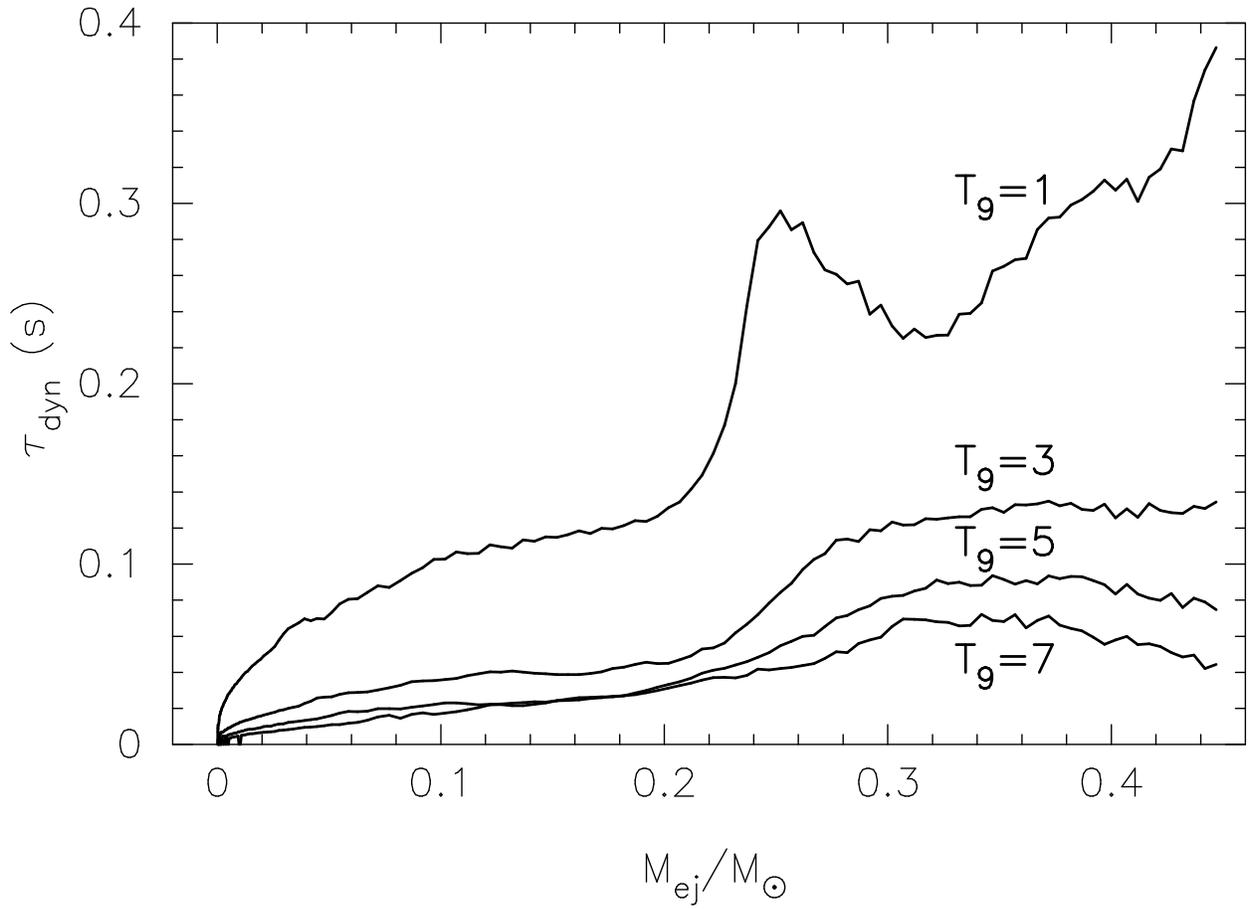} 
\caption{The dynamical timescales of the
outgoing material as functions of $M_{\rm ej}$ at $T_9 = 1, 3, 5$, and 7
in model~Q6.}
\end{figure}

\clearpage

\begin{figure}
\epsscale{1.0} 
\plotone{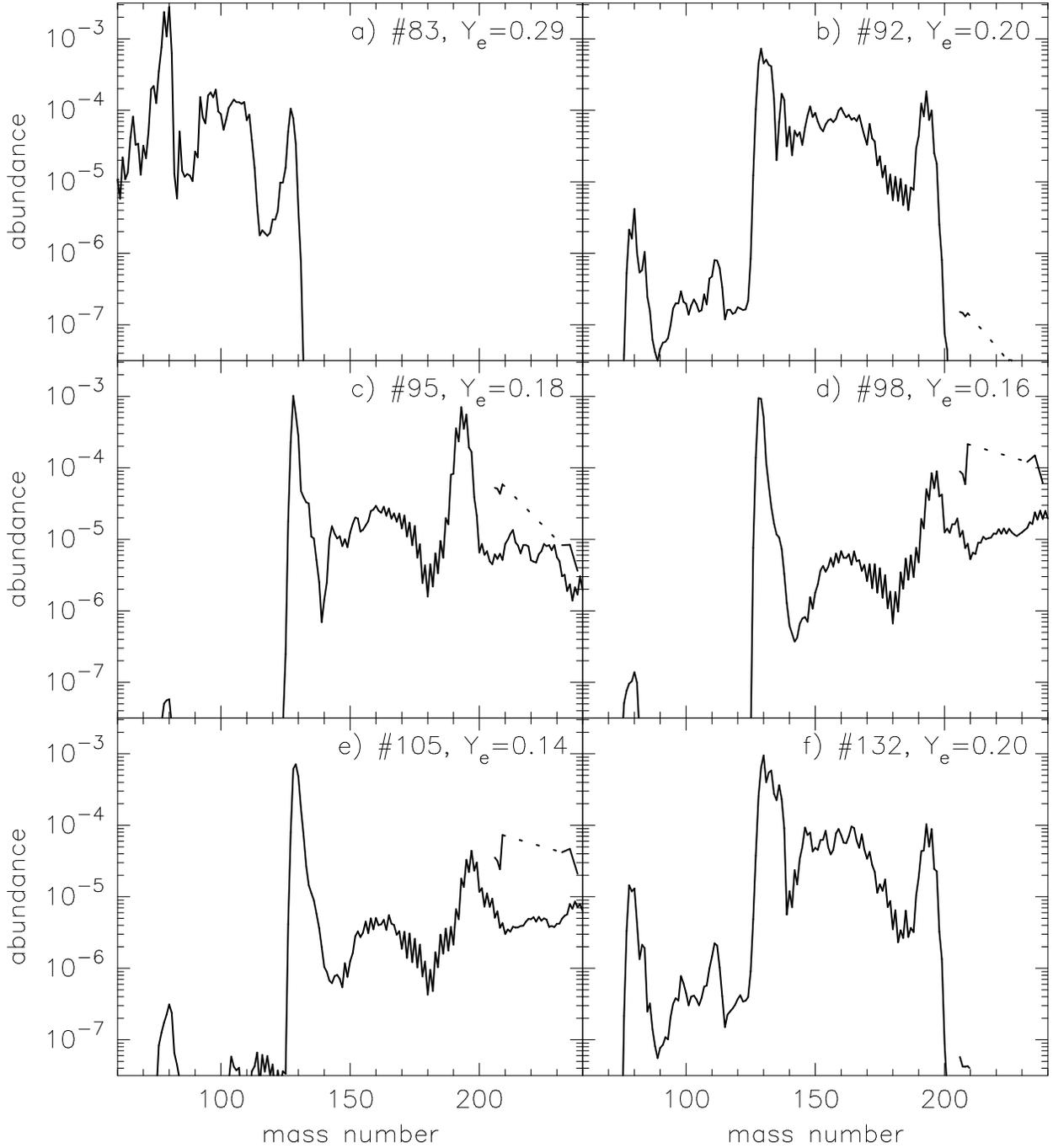} 
\caption{ Final abundances as a function
of mass number from $r$-process calculations for trajectories (a) 83,
(b) 92, (c) 95, (d) 98, (e) 105, and (f) 132 in Table~2.  }
\end{figure}

\clearpage

\begin{figure}
\epsscale{1.0} 
\plotone{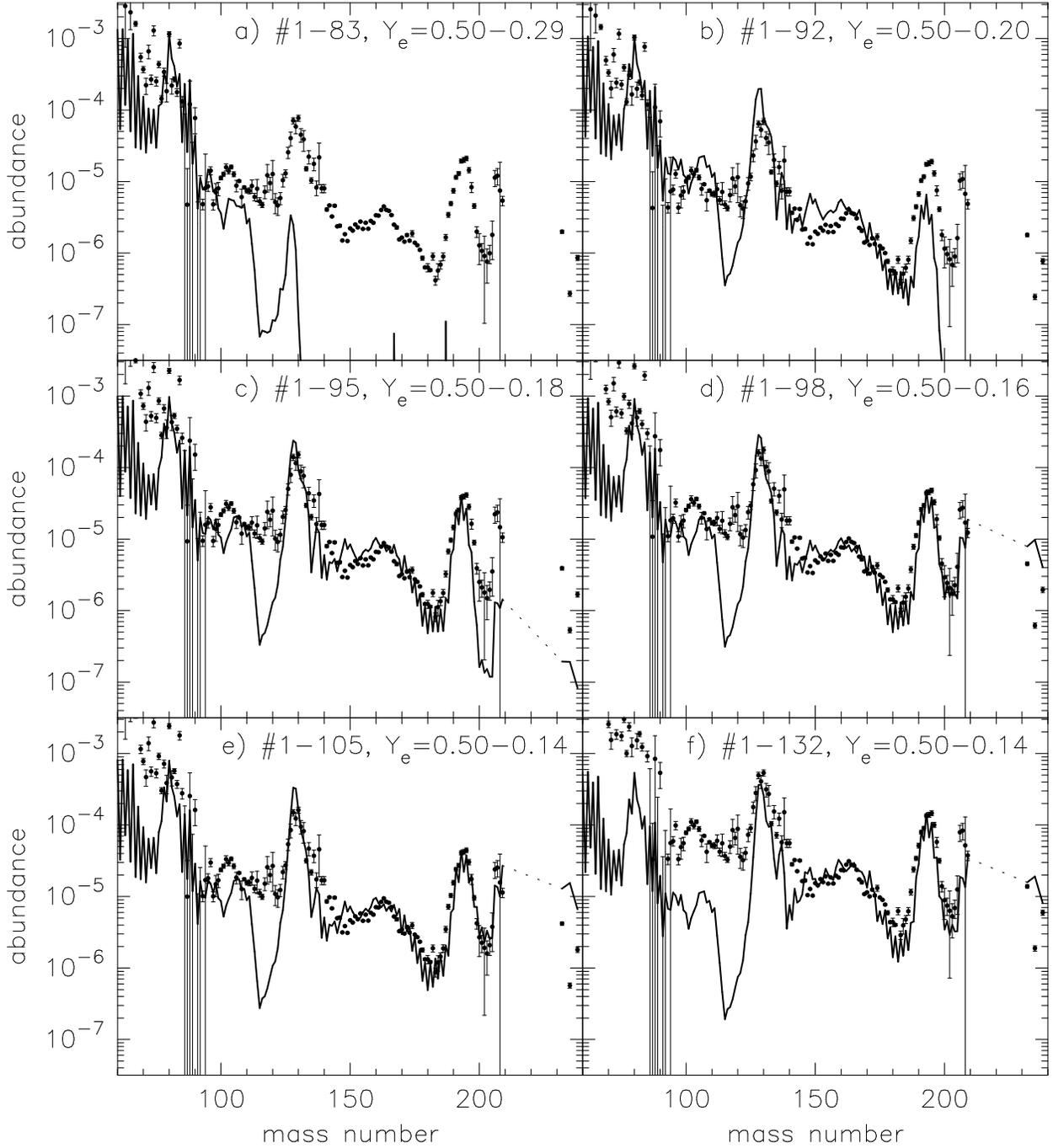} 
\caption{ Final mass-averaged
$r$-process abundances (line) as a function of mass number obtained from
the ejected zones in (a) models~Q6a, (b) Q6b, (c) Q6c, (d) Q6d, (e) Q6e,
and (f) Q6f (see Table~2). These are compared with the solar $r$-process
abundances (points) of \citet{Kapp89}, which is scaled to match the
height of the first peak ($A = 80$) for (a)-(b) and the third peak ($A =
195$) for (c)-(f).}
\end{figure}

\clearpage

\begin{figure}
\epsscale{1.0} 
\plotone{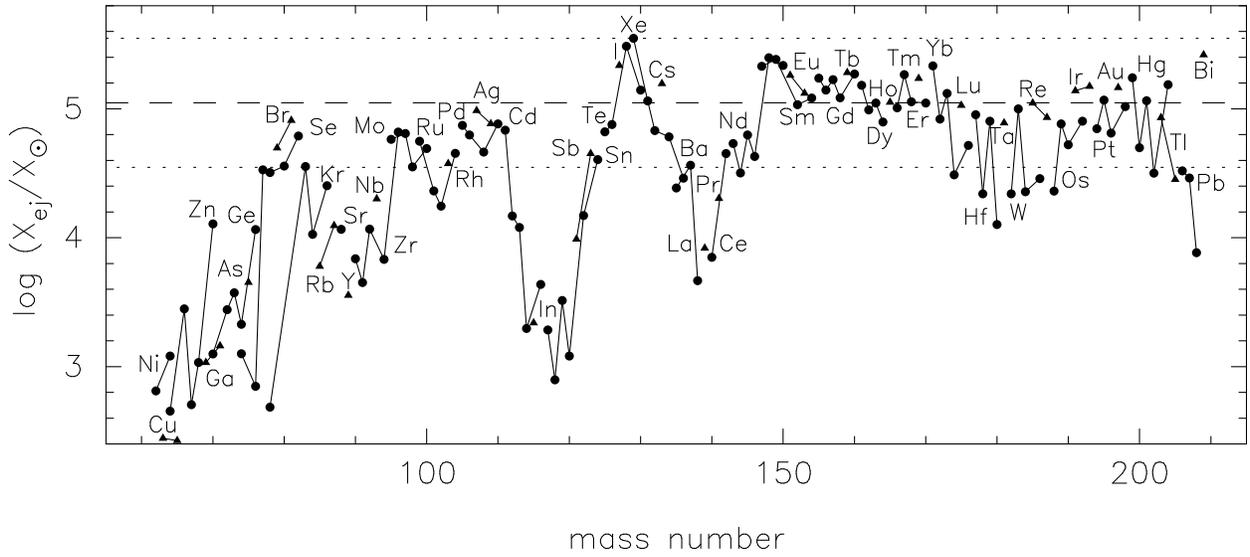} 
\caption{ Mass-averaged production
factors in model~Q6e (see Table~2). Isotopes of a given element are
connected by lines. Elements with even and odd atomic numbers are
denoted by points and triangles, respectively. The dotted lines indicate
a normalization band (see text), with its median value (dashed line).  }
\end{figure}

\clearpage

\begin{figure}
\epsscale{1.0} 
\plotone{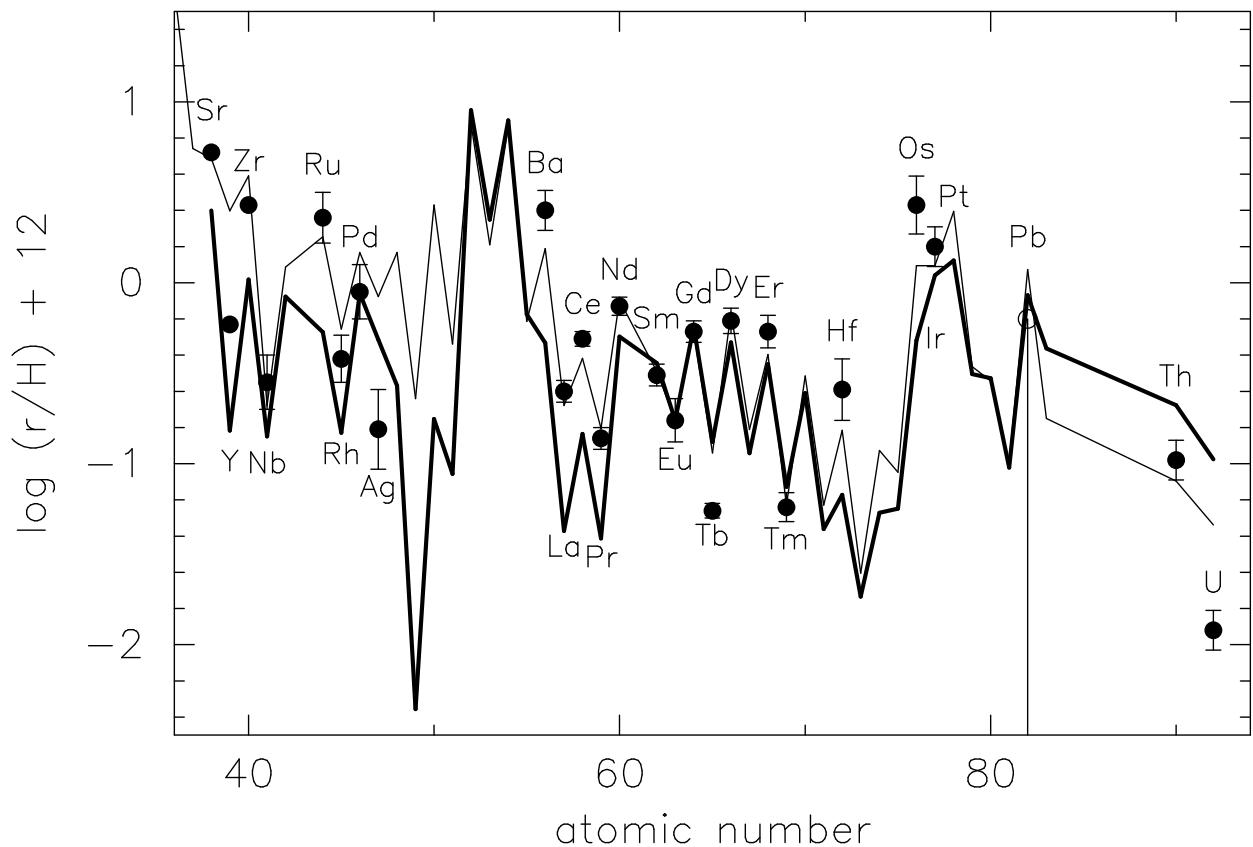} 
\caption{ Comparison of the
mass-integrated yields (thick line) for model~Q6e, scaled at Eu ($Z =
63$), with the abundance pattern of CS~31082-001 (filled circles, with
observational error bars), as a function of atomic number. For Pb, the
observed upper limit is shown by the open circle. The scaled solar
$r$-process pattern is shown by the thin line. }
\end{figure}

\clearpage

\begin{figure}
\epsscale{1.0} 
\plotone{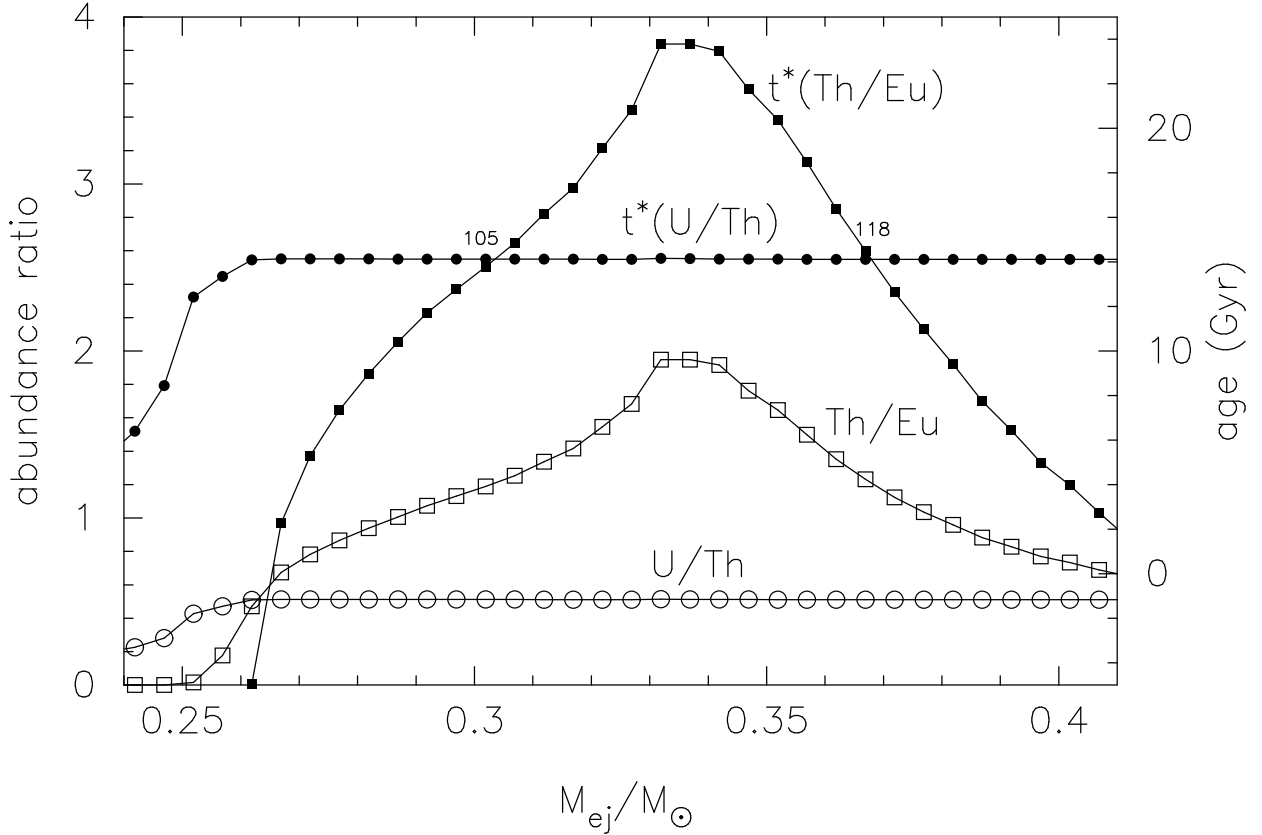} 
\caption{Mass-integrated abundance
ratios Th/Eu (open squares) and U/Th (open circles) from the surface of
the core to the mass point $M_{\rm ej}$ in model~Q6. The surface of the
O-Ne-Mg core is at mass coordinate zero. Ages of CS~31082-001 $t^* ({\rm
Th/Eu})$ (filled squares) and $t^* ({\rm U/Th})$ (filled circles)
inferred by these ratios are also shown. The lines of these ages have
intersections near the mass points 105 and 118.  }
\end{figure}


\clearpage

\begin{thebibliography}{}
\bibitem[Anders \& Grevesse(1989)]{Ande89}
 Anders, E., \& Grevesse, N. 1989, \gca, 53, 197
\bibitem[Bailyn \& Grindlay(1990)]{Bail90}
 Bailyn, C. D. \& Grindlay, J. E. 1990, \apj, 353, 159
\bibitem[Baron, Cooperstein, \& Kahana(1987)]{Baro87a}
 Baron, E., Cooperstein, J., \& Kahana, S. 1987, \apj, 320, 300
\bibitem[Baron et al.(1987)]{Baro87b}
 Baron, E., Cooperstein, J., Kahana, S., \& Nomoto, K. 1987, \apj, 320, 304
\bibitem[Baron \& Cooperstein(1990)]{Baro90}
 Baron, E. \& Cooperstein, J. 1990, \apj, 353, 597
\bibitem[Bowers \& Wilson(1982)]{Bowe82}
 Bowers, R. \& Wilson, J. R. 1982, \apj, 263, 366
\bibitem[Bowers \& Wilson(1991)]{Bowe91}
 Bowers, R. \& Wilson, J. R. 1991, Numerical Modeling in Applied Physics
 and Astrophysics (Jones and Bartlett)
\bibitem[Bruenn (1989a)]{Brue89a}
 Bruenn, S. W. 1989, \apj, 340, 955
\bibitem[Bruenn (1989b)]{Brue89b}
 Bruenn, S. W. 1989, \apj, 341, 385
\bibitem[Burrows \& Lattimer(1983)]{Burr83}
 Burrows, A. \& Lattimer, J. M. 1983, \apj, 270, 735
\bibitem[Burrows \& Lattimer(1985)]{Burr85}
 Burrows, A. \& Lattimer, J. M. 1985, \apjl, 299, L19
\bibitem[Cappellaro et al.(1997)]{Capp97}
 Cappellaro, E., Turatto, M., Tsvetkov, D. Yu., Bartunov, O. S., Pollas,
 C., Evans, R., Hamuy, M. 1997, \aap, 322, 431
\bibitem[Cardall \& Fuller(1997)]{Card97}
 Cardall, C. Y. \& Fuller, G. M. 1997, \apjl, 486, L111
\bibitem[Cayrel et al.(2001)]{Cayr01}
 Cayrel, R., Hill, V., Beers, T. C., Barbuy, B., Spite, M., Spite, F.,
 Plez, B., Andersen, J., Bonifacio, P., Francois, P., Molaro, P.,
 Nordstrom, B., \& Primas, F. 2001, \nat, 409, 691
\bibitem[Christlieb et al.(2002)]{Chri02}
 Christlieb, N., Bessell, M. S., Beers, T. C., Gustafsson, B., Korn, A.,
 Barklem, P. S., Karlsson, T., Mizuno-Wiedner, M., \& Rossi, S.
 2002, \nat, 419, 904
\bibitem[Cowan, Thielemann, \& Truran(1991)]{Cowa91}
 Cowan, J. J., Thielemann, F. -K., \& Truran, J. W. 1991, \physrep, 208, 267
\bibitem[Cowan et al.(2002)]{Cowa02}
 Cowan, J. J., Sneden, C., Burles, S., Ivans, I. I.,
 Beers, T. C., Truran, J. W., Lawler, J. E., Primas,
 F., Fuller, G. M., Pfeiffer, B., \& Kratz, K. -L.
 2002, \apj, 572, 861
\bibitem[Fryer et al.(1999)]{Frye99}
 Fryer, C., Benz, W., Herant, M., \& Colgate, S. A. 1999, \apj, 516, 892
\bibitem[Fryer \& Heger(2000)]{Frye00}
 Fryer, C. \& Heger, A. 2000, \apj, 541, 1033
\bibitem[Goriely(1997)]{Gori97}
 Goriely, S. 1997, \aap, 325, 414
\bibitem[Goriely \& Arnould(2001)]{Gori01}
 Goriely, S. \& Arnould, M. 2001, \aap, 379, 1113
\bibitem[Hilf et al.(1976)]{Hilf76}
 Hilf, E. R., von Groote, H., \& Takahashi, K. 1976, in Proc. Third
 International Conference on Nuclei Far from Stability (Geneva: CERN), 142
\bibitem[Hachisu et al.(1990)]{Hach90}
 Hachisu, I., Matsuda, T., Nomoto, K., \& Shigeyama, T. 1990, \apjl, 358, L57
\bibitem[Herant \& Woosley(1994)]{Hera94}
 Herant, M. \& Woosley, S. E. 1994, \apj, 425, 814
\bibitem[Hill et al.(2002)]{Hill02}
 Hill, V., Plez, B., Cayrel, R., Beers, T.C., Nordstr\"om, B., 
 Andersen, J., Spite, M., Spite, F., Barbuy, B., Bonifacio, P., 
 Depagne, E., Fran\c{c}ois, P., Molaro, P., \& Primas, F.
 2002, \aap, 387, 560
\bibitem[Hillebrandt, Takahashi, \& Kodama(1976)]{Hill76}
 Hillebrandt, W., Takahashi, K., \& Kodama, T. 1976, \aap, 52, 63
\bibitem[Hillebrandt, Nomoto, \& Wolff(1984)]{Hill84}
 Hillebrandt, W., Nomoto, K., \& Wolff, G. 1984, \aap, 133, 175
\bibitem[Hoffman et al.(1997)]{Hoff97}
 Hoffman, R. D., Woosley, S. E., \& Qian, Y. -Z. 1997, \apj, 482, 951
\bibitem[Ishimaru \& Wanajo(1999)]{Ishi99}
 Ishimaru, Y. \& Wanajo, S. 1999, \apjl, 511, L33
\bibitem[Ishimaru \& Wanajo(2000)]{Ishi00}
 Ishimaru, Y. \& Wanajo, S. 2000, in First Stars, ed. A. Weiss,
 T. Abel,\& V. Hill (Berlin: Springer), 189
\bibitem[Johnson \& Bolte(2002)]{John02}
 Johnson, J. A. \& Bolte, M. 2002, \apj, 579, 616
\bibitem[K\"appeler et al.(1989)]{Kapp89}
K\"appeler, F., Beer, H., \& Wisshak, K. 1989, Rep. Prog. Phys., 52, 945
\bibitem[Kifonidis et al.(2003)]{Kifo03}
 Kifonidis, K., Plewa, T., Janka, H.-Th. \& E. Mueller 2003, \aap,
 submitted (astro-ph/0302239)
\bibitem[Langanke \& Martinez-Pinedo(2000)]{Lang00}
 Langanke, K. \& Martinez-Pinedo, G. 2000, \nphysa, 673, 481
\bibitem[Mayle \& Wilson(1988)]{Mayl88}
 Mayle, R. \& Wilson, J. R. 1988, \apj, 334, 909
\bibitem[Meyer, McLaughlin, \& Fuller(1998)]{Meye98}
 Meyer, B. S., McLaughlin, G. C., \& Fuller, G. M. 1998, \prc, 58, 3696
\bibitem[Miyaji et al.(1980)]{Miya80}
 Miyaji, S., Nomoto, K., Yokoi, K., \& Sugimoto, D. 1980, \pasj, 32, 303
\bibitem[Miyaji \& Nomoto(1987)]{Miya87}
 Miyaji, S. \& Nomoto, K. 1987, \apj, 318, 307
\bibitem[Nomoto et al.(1982)]{Nomo82}
 Nomoto, K., Sparks, W. M., Fesen, R. A., Gull, T. R., Miyaji, S.,
 \& Sugimoto, D. 1982, \nat, 299, 803
\bibitem[Nomoto(1984)]{Nomo84}
 Nomoto, K. 1984, \apj, 277, 791
\bibitem[Nomoto(1987)]{Nomo87}
 Nomoto, K. 1987, \apj, 322, 206
\bibitem[Nomoto \& Kondo(1991)]{Nomo91}
 Nomoto, K. \& Kondo, Y. 1991, \apjl, 367, L19
\bibitem[Otsuki et al.(2000)]{Otsu00}
 Otsuki, K., Tagoshi, H., Kajino, T., \& Wanajo, S. 2000, \apj, 533, 424
\bibitem[Qian \& Woosley(1996)]{Qian96}
 Qian, Y. -Z. \& Woosley, S. E. 1996, \apj, 471, 331
\bibitem[Qian et al.(1997)]{Qian97}
 Qian, Y. -Z., Haxton, W. C., Langanke, K., \& Vogel, P. 1997, \prc, 55, 1532
\bibitem[Qian, Vogel, \& Wasserburg(1998)]{Qian98}
 Qian, Y. -Z., Vogel, P., \& Wasserburg, G. J. 1998, \apj, 506, 868
\bibitem[Qian \& Wasserburg(2001)]{Qian01}
 Qian, Y. -Z. \& Wasserburg, G. J. 2001, \apj, 559, 925
\bibitem[Qian \& Wasserburg(2002)]{Qian02}
 Qian, Y. -Z. \& Wasserburg, G. J. 2002, \apj, 567, 515
\bibitem[Qian \& Wasserburg(2003)]{Qian03}
 Qian, Y. -Z. \& Wasserburg, G. J. 2003, \apj, in press (astro-ph/0301461)
\bibitem[Sato(1974)]{Sato74}
 Sato, K. 1974, Prog. Theor. Phys., 51, 726
\bibitem[Schatz et al.(2002)]{Scha02}
 Schatz, H., Toenjes, R., Kratz, K.-L., Pfeiffer, B., Beers, T.C., Cowan, J.J.,
\& Hill, V. 2002, \apj, 579, 626
\bibitem[Seeger \& Schramm(1970)]{Seeg70}
 Seeger, P. A. \& Schramm, D. N. 1970, \apjl, 160, L157
\bibitem[Schramm(1973)]{Schr73}
 Schramm, D. N. 1973, \apj, 185, 293
\bibitem[Shen et al.(1998)]{Shen98}
 Shen, H., Toki, H., Oyamatsu, K., \& Sumiyoshi, K. 1998, \nphysa, 637, 435
\bibitem[Sneden et al.(1996)]{Sned96}
 Sneden, C., McWilliam, A., Preston, G. W., Cowan, J. J., Burris, D. L.,
 \& Armosky, B. J. 1996, \apj, 467, 819
\bibitem[Sneden et al.(2000)]{Sned00}
 Sneden, C., Cowan, J. J., Ivans, I. I., Fuller, G. M., Burles, S.,
 Beers, T. C., Lawler, J. E. 2000, \apjl, 533, L139
\bibitem[Sneden \& Cowan(2003)]{Sned03a}
 Sneden, C. \& Cowan, J. J. 2003, Science, 299, 70
\bibitem[Sneden et al.(2003)]{Sned03b}
 Sneden, C., Cowan, J. J., Lawler, J. E., Ivans,  I. I., Burles, S.,
 Beers, T. C., Primas, F., Hill, V., Truran, J. W., Fuller, G. M.,
 Pfeiffer, B., \& Kratz, K. -L.
 2003, \apj, in press (astro-ph/0303542)
\bibitem[Sumiyoshi et al.(2001)]{Sumi01}
 Sumiyoshi, K., Terasawa, M., Mathews, G. J., Kajino, T., Yamada, S., \&
 Suzuki, H. 2001, \apj, 562, 880
\bibitem[Surman \& Engel(2001)]{Surm01}
 Surman, R. \& Engel 2001, \prc, 64, 35801
\bibitem[Thompson, Burrows, \& Meyer(2001)]{Thom01}
 Thompson, T. A., Burrows, A., \& Meyer, B. S. 2001, \apj, 562, 887
\bibitem[Thompson, Burrows, \& Pinto(2003)]{Thom03}
 Thompson, T. A., Burrows, A., \& Pinto, P. A. 2003, \apj, in press
 (astro-ph/0211194)
\bibitem[Umeda \& Nomoto(2002)]{Umed02}
 Umeda, H. \& Nomoto, K. 2002, \apj, 565, 385
\bibitem[Umeda \& Nomoto(2003)]{Umed03}
 Umeda, H. \& Nomoto, K. 2003, \nat, in press (astro-ph/0301315)
\bibitem[Utsunomiya et al.(2001)]{Utsu01}
 Utsunomiya, H., Yonezawa, Y., Akimune, H., Yamagata, T., Ohta, M., 
 Fujishiro, M., Toyokawa, H., \& Phgaki, H.  2001, \prc, 63, 18801
\bibitem[Wallerstein et al.(1995)]{Wall95}
 Wallerstein, G., Vanture, A. D., Jenkins, E. B., \& Fuller, G. M. 1995,
 \apj, 449, 688
\bibitem[Wanajo et al.(2001)]{Wana01}
 Wanajo, S., Kajino, T., Mathews, G. J., \& Otsuki, K. 2001, \apj, 554, 578
\bibitem[Wanajo et al.(2002)]{Wana02}
 Wanajo, S., Itoh, N., Ishimaru, Y., Nozawa, S., \& Beers, T. C. 2002,
 \apj, 577, 853
\bibitem[Wasserburg \& Qian(2000)]{Wass00}
 Wasserburg, G. J. \& Qian, Y. -Z., 2000 \apjl, 529, L21
\bibitem[Wheeler, Cowan, \& Hillebrandt(1998)]{Whee98}
 Wheeler, J. C., Cowan, J. J., \& Hillebrandt, W. 1998, \apjl, 493, L101
\bibitem[Woosley \& Baron(1992)]{Woos92a}
 Woosley, S. E. \& Baron, E. 1992, \apj, 391, 228
\bibitem[Woosley \& Hoffman(1992)]{Woos92b}
 Woosley, S. E. \& Hoffman, R. D. 1992, \apj, 395, 202
\bibitem[Woosley et al.(1994)]{Woos94}
 Woosley, S. E., Wilson, J. R., Mathews, G. J., Hoffman, R. D., \&
 Meyer, B. S. 1994, \apj, 433, 229
\bibitem[Woosley \& Weaver(1995)]{Woos95}
 Woosley, S. E. \& Weaver, T. A. 1995, \apjs, 101, 181
\bibitem[Yamada \& Sato(1994)]{Yama94}
 Yamada, S. \& Sato, K. 1994, \apj, 434, 268
\end{thebibliography}
\end{document}